\documentclass[12pt]{article}
\usepackage{amssymb,graphicx}
\usepackage{epstopdf}

\begin{document}

\newcommand{\sect}[1]{\setcounter{equation}{0}\section{#1}}
\renewcommand{\theequation}{\thesection.\arabic{equation}}
\newcommand{\be}{\begin{equation}}
\newcommand{\ee}{\end{equation}}
\newcommand{\bea}{\begin{eqnarray}}
\newcommand{\eea}{\end{eqnarray}}
\newcommand{\eps}{\epsilon}
\newcommand{\om}{\omega}
\newcommand{\vph}{\varphi}
\newcommand{\sig}{\sigma}
\newcommand{\CC}{\mbox{${\mathbb C}$}}
\newcommand{\RR}{\mbox{${\mathbb R}$}}
\newcommand{\QQ}{\mbox{${\mathbb Q}$}}
\newcommand{\ZZ}{\mbox{${\mathbb Z}$}}
\newcommand{\NN}{\mbox{${\mathbb N}$}}
\newcommand{\1}{\mbox{\hspace{.0em}1\hspace{-.24em}I}}
\newcommand{\II}{\mbox{${\mathbb I}$}}
\newcommand{\prt}{\partial}
\newcommand{\und}[1]{\underline{#1}}
\newcommand{\wh}[1]{\widehat{#1}}
\newcommand{\wt}[1]{\widetilde{#1}}
\newcommand{\mb}[1]{\ \mbox{\ #1\ }\ }
\newcommand{\half}{\frac{1}{2}}
\newcommand{\noin}{\not\!\in}
\newcommand{\rhotimes}{\mbox{\raisebox{-1.2ex}{$\stackrel{\displaystyle\otimes}
{\mbox{\scriptsize{$\rho$}}}$}}}
\newcommand{\bin}[2]{{\left( {#1 \atop #2} \right)}}
\newcommand{\ri}{{\rm i}}
\newcommand{\A}{{\cal A}}
\newcommand{\B}{{\cal B}}
\newcommand{\C}{{\cal C}}
\newcommand{\F}{{\cal F}}
\newcommand{\E}{{\cal E}}
\newcommand{\cP}{{\cal P}}
\newcommand{\R}{{\cal R}}
\newcommand{\T}{{\cal T}}
\newcommand{\W}{{\cal W}}
\newcommand{\cS}{{\cal S}}
\newcommand{\bS}{{\bf S}}
\newcommand{\cL}{{\cal L}}
\newcommand{\hlp}{{\RR}_+}
\newcommand{\hlm}{{\RR}_-}
\newcommand{\Hil}{{\cal H}}
\newcommand{\D}{{\cal D}}
\newcommand{\G}{{\cal G}}
\newcommand{\alg}{\C} 
\newcommand{\rep}{\F(\A)}
\newcommand{\trep}{\G_\beta(\A)}
\newcommand{\form}{\langle \, \cdot \, , \, \cdot \, \rangle }
\newcommand{\e}{{\rm e}}
\newcommand{\by}{{\bf y}}
\newcommand{\bp}{{\bf p}}
\newcommand{\LL}{\mbox{${\mathbb L}$}}
\newcommand{\Rp}{{R^+_{\, \, \, \, }}}
\newcommand{\Rm}{{R^-_{\, \, \, \, }}}
\newcommand{\Rpm}{{R^\pm_{\, \, \, \, }}}
\newcommand{\Tp}{{T^+_{\, \, \, \, }}}
\newcommand{\Tm}{{T^-_{\, \, \, \, }}}
\newcommand{\Tpm}{{T^\pm_{\, \, \, \, }}}
\newcommand{\baral}{\bar{\alpha}}
\newcommand{\barbt}{\bar{\beta}}
\newcommand{\supp}{{\rm supp}\, }
\newcommand{\EE}{\mbox{${\mathbb E}$}}
\newcommand{\JJ}{\mbox{${\mathbb J}$}}
\newcommand{\MM}{\mbox{${\mathbb M}$}}
\newcommand{\ct}{{\cal T}}
\newcommand{\ph}{\varphi}
\newcommand{\phd}{\widetilde{\varphi}}
\newcommand{\phl}{\varphi_{{}_L}}
\newcommand{\phr}{\varphi_{{}_R}}
\newcommand{\phpl}{\varphi_{{}_{+L}}}
\newcommand{\phpr}{\varphi_{{}_{+R}}}
\newcommand{\phml}{\varphi_{{}_{-L}}}
\newcommand{\phmr}{\varphi_{{}_{-R}}}
\newcommand{\phpml}{\varphi_{{}_{\pm L}}}
\newcommand{\phpmr}{\varphi_{{}_{\pm R}}}
\newcommand{\Ei}{\rm Ei}

\newtheorem{theo}{Theorem}[section]
\newtheorem{coro}[theo]{Corollary}
\newtheorem{prop}[theo]{Proposition}
\newtheorem{defi}[theo]{Definition}
\newtheorem{conj}[theo]{Conjecture}
\newtheorem{lem}[theo]{Lemma}
\newcommand{\prf}{\underline{\it Proof.}\ }
\newcommand{\finprf}{\null \hfill {\rule{5pt}{5pt}}\\[2.1ex]\indent}

\pagestyle{empty}
\rightline{May 2006}

\vfill

\begin{center}
{\Large\bf Quantum Fields on Star Graphs}
\\[2.1em]

\bigskip

{\large
B. Bellazzini$^{a}$\footnote{b.bellazzini@sns.it} and  
M. Mintchev$^{b}$\footnote{mintchev@df.unipi.it}}\\

\null

\noindent 

{\it 
$^a$ INFN and Scuola Normale Superiore, Piazza dei Cavalieri 7, 
56127 Pisa,  Italy\\[2.1ex]
$^b$ INFN and Dipartimento di Fisica, Universit\`a di
      Pisa, Largo Pontecorvo 3, 56127 Pisa, Italy }
\vfill

\end{center}

\begin{abstract} 

We construct canonical quantum fields which propagate on a star graph 
modeling a quantum wire. The construction uses a deformation 
of the algebra of canonical commutation relations, encoding the 
interaction in the vertex of the graph. We discuss in this framework the 
Casimir effect and derive the correction to the Stefan-Boltzmann law 
induced by the vertex interaction. We also generalize the algebraic setting in 
order to cover systems with integrable bulk interactions and solve the quantum 
non-linear Schr\"odinger model on a star graph.

\end{abstract}
\bigskip 
\medskip 
\bigskip

\vfill
\rightline{IFUP-TH 10/2006}
\rightline{\tt hep-th/0605036}
\newpage
\pagestyle{plain}
\setcounter{page}{1}

\sect{Introduction}

The physics of thin three-dimensional graph-like structures \cite{D}, two of 
whose dimensions being of  
the order of a few nanometers, attracts recently much attention. 
Due to the impressive progress in nanotechnology 
such devices, called quantum wires, can be constructed 
and tested nowadays in laboratory. At the theoretical side, the quantum behavior of 
these essentially one-dimensional systems poses interesting 
mathematical and physical problems, 
which have been addressed in the past decade by various authors 
\cite{E}-\cite{Fulling:2005js}. 
In the present paper we pursue further the study of quantum wires,
developing field theory on such structures. For this purpose we 
adapt the algebraic approach developed in 
\cite{Liguori:1996xr}-\cite{Mintchev:2003ue} for dealing with 
systems with boundaries and point-like defects. This approach 
translates the boundary value problem at hand in 
algebraic terms and offers an efficient framework for the construction of 
quantum fields. 
\vskip 0.5truecm 
\setlength{\unitlength}{1,25mm}
\begin{picture}(20,20)(-25,20) 
\put(25.2,0.7){\makebox(20,20)[t]{$\bullet$}}
\put(28.5,1){\makebox(20,20)[t]{$V$}}
\put(42,11){\makebox(18,22)[t]{$E_1$}}
\put(33,17){\makebox(20,20)[t]{$E_2$}}
\put(9,3.5){\makebox(20,20)[t]{$E_i$}}
\put(34.5,-12){\makebox(20,20)[t]{$E_n$}}
\thicklines 
\put(35,20){\line(1,1){16}}
\put(35,20){\line(-1,0){19}}
\put(35,20){\line(1,-1){13}}
\put(35,20){\line(1,3){6}}
\put(20,3){\makebox(20,20)[t]{$.$}}
\put(20.9,5){\makebox(20,20)[t]{$.$}}
\put(23.8,6.6){\makebox(20,20)[t]{$.$}}
\put(20,-3){\makebox(20,20)[t]{$.$}}
\put(20.9,-5){\makebox(20,20)[t]{$.$}}
\put(23.8,-6.6){\makebox(20,20)[t]{$.$}}
\end{picture}
\vskip 2.5 truecm
\centerline{Figure 1: A star graph $G$ with $n$ edges.}
\bigskip 

We focus in this paper on quantum wires of the form of a star graph $G$  
shown in Fig. 1, where $V$ stands for the vertex of the graph and 
$E_i$ (with $i=1,...,n$) denote its edges. Being the fundamental building 
blocks for general graphs, the star graphs play a distinguished role 
\cite{Kostrykin:1998gz, KS} in the subject. The first question, we address 
in section 2, is the explicit construction of a canonical 
scalar quantum field, freely 
propagating in the bulk $B\equiv G \setminus V$ of the graph $G$. 
The vertex is considered in our approach as a defect (boundary in the case 
$n=1$) interacting with the field and characterized by reflection ($n\geq 1$) 
and transmission ($n\geq 2$) coefficients. We consider the case without dissipation and 
treat both relativistic and non-relativistic dispersion relations. 
In section 3 we derive the correlation functions with respect to two 
different cyclic states - the Fock vacuum and a Gibbs state at finite 
(inverse) temperature $\beta$. Some basic physical observables are analyzed in 
section 4. In particular, we compute there the Casimir energy and the 
correction to the Stefan-Boltzmann law generated by the interaction 
localized in the vertex of the graph. Inspired by factorized scattering in 1+1 dimensional 
integrable systems, we introduce in section 5 a 
two-body interaction in the bulk $B$. The system treated here is the non-linear 
Schr\"odinger model (non-relativistic $\psi^4$ theory) on a star graph. 
We establish the exact operator solution of this model. Section 6 contains 
our conclusions and some ideas for future developments. 

\bigskip 

\sect{Quantum fields on a star graph} 

Consider a star graph $G$. Each point $P$ in the bulk $B$ of $G$ can be 
parametrized by a pair $(i,x)$ where $i=1,...,n$ indicates the edge and 
$x>0$ the distance of $P$ from the vertex $V$ along the edge. 
We will see below that this choice of coordinates 
ensures a uniform treatment of all edges as far as the
reflection  and transmission by the vertex $V$ is concerned. 
The embedding of $G$ and the relative position of 
its edges $E_i$ in the ambient space are irrelevant in what follows.

\subsection{Non-relativistic dispersion}

We start with the non-relativistic complex scalar field $\psi (t,x,i)$ 
satisfying   
\be
\left (\ri \prt_t +\frac{1}{2m} \prt_x^2\right )\psi_i (t,x) = 0\, , 
\qquad t\in \RR \, ,\quad  x>0 \, , \quad i=1,...,n\, , 
\label{eqm}
\ee
with standard initial conditions given by the equal-time 
canonical commutation relations 
\be
[\psi_{i_1} (0,x_1)\, ,\, \psi_{i_2} (0,x_2)] = 
[\psi^{*i_1} (0,x_1)\, ,\, \psi^{*i_2} (0,x_2)]=0  \, , 
\label{initial1} 
\ee
\be
[\psi_{i_1} (0,x_1)\, ,\, \psi^{*i_2} (0,x_2)] =  
\delta_{i_1}^{i_2}\, \delta (x_1-x_2)  \, ,  
\label{initial2}
\ee 
where $*$ denotes Hermitian conjugation. Since 
eqs. (\ref{eqm}-\ref{initial2}) hold only in the bulk $B$, they do not fix uniquely 
the solution. For this purpose one needs some  
boundary conditions in the vertex $V$. Setting 
\be 
\psi_i (t,0)=\lim_{x\to 0\atop x>0}\psi_i (t,x)\, , \qquad 
(\prt_x\psi_i ) (t,0)=\lim_{x\to 0\atop x>0}(\prt_x\psi_i ) (t,x)\, , 
\label{deff}
\ee 
we require \cite{Kostrykin:1998gz} that\footnote{Summation over repeated upper 
and lower indices is understood, unless otherwise stated.}  
\be 
A_i^j \psi_j (t,0) + B_i^j (\prt_x\psi_j ) (t,0) = 0\, , \quad \forall \, t\in \RR\, , \quad i=1,...,n\, , 
\label{bc} 
\ee 
$A$ and $B$ being two complex $n\times n$ matrices. 
It has been shown in \cite{Kostrykin:1998gz}, 
that (\ref{bc}) uniquely defines a self-adjoint extension of the operator 
$-\prt_x^2$ on $G$, provided that the composite matrix $(A,B)$ has a 
rank $n$ and 
\be 
A\, B^* - B\, A^* = 0 \, . 
\label{cond1} 
\ee
Moreover, this extension 
admits a non-trivial scattering matrix, reading in momentum space 
\be 
S(k) = -(A+\ri kB )^{-1} (A-\ri kB) \, , 
\label{S1}
\ee
which, in view of (\ref{cond1}), can be written also in the form 
\be 
S(k) = -(A^* -\ri k B^*)(A A^*+k^2 B B^*)^{-1} (A -\ri k B)\, . 
\label{S2}
\ee
The entries of $S$ have a simple physical interpretation: the diagonal 
element $S_i^i(k)$ is the reflection amplitude on the edge $E_i$, 
whereas $S_i^j(k)$ with $i\not=j$ is the transmission amplitude 
from $E_i$ to $E_j$. From (\ref{S1}) and (\ref{S2}) one can easily deduce that 
$S$ is unitary
\be 
S(k)^*=S(k)^{-1} \, , 
\label{unit1}
\ee 
and satisfies Hermitian analyticity 
\be 
S(k)^*=S(-k)\, .  
\label{Ha}
\ee
Combining (\ref{unit1}) and (\ref{Ha}) one gets 
\be 
S(k)\, S(-k) = \II_n \, , 
\label{unit2} 
\ee
where $ \II_n$ is the $n\times n$ identity matrix. The property (\ref{unit2}) is 
essential in what follows. 

Turning back to the quantum field $\psi$, we introduce 
the wave functions \cite{Kostrykin:1998gz}
\be 
\chi_i^j(x;k) \equiv \e^{\ri kx} \delta_i^j + 
S_{i}^{j} (-k)\e^{-\ri kx} \, , \qquad k>0\, ,   
\label{chi}
\ee 
which are orthogonal (see the appendix) 
\be 
\int_{0}^{+\infty} dx\, \chi_{\, i}^{*l}(x;k) \chi_l^j(x;p) = \delta_i^j\, 2\pi\, \delta (k-p) \, .  
\label{chiort}
\ee 
If in addition  
\be
{\check S}_i^j(x) \equiv 
\int_{-\infty}^{+\infty} \frac{dk}{2\pi } \e^{\ri kx} S_i^j(k) = 0\, , 
\qquad x>0\, , 
\label{compl1}
\ee 
then $\chi$ form a complete set, i.e. 
\be 
\int_{0}^{+\infty} \frac{dk}{2\pi}\, \chi_{\, i}^{*l}(x;k) \chi_l^j(y;k) = \delta_i^j\, \delta (x-y) \, .  
\label{compl2}
\ee
At this point we define 
\be 
\psi_i (t,x) = \int_{0}^{+\infty} 
\frac{dk}{2\pi }\e^{-\ri \omega (k)t} \chi_i^j(x;k) a_j (k)\, , 
\label{phi1}
\ee
\be 
\psi^{*i} (t,x) = \int_{0}^{+\infty} 
\frac{dk}{2\pi }\e^{\ri \omega (k)t} a^{*j}(k) \chi_{\, j}^{*i}(x;k)\, , 
\label{phi2}
\ee 
where $\omega(k)$ is the dispersion relation 
\be 
\omega(k) = \frac {k^2}{2m} \, , \qquad m>0\, , 
\label{disp1}
\ee 
and $\{a_i(k),\, a^{* i}(k)\, :\, k>0\}$ generate an associative algebra $\A_+$ 
with identity element $\bf 1$ and satisfy the canonical commutation relations  
\bea
&a_{i_1}(k_1)\, a_{i_2}(k_2) -  a_{i_2}(k_2)\, a_{i_1}(k_1) = 0\,  ,
\label{ccr1} \\
&a^{\ast i_1}(k_1)\, a^{\ast i_2}(k_2) - a^{\ast i_2}(k_2)\,
a^{\ast i_1}(k_1) = 0\,  ,
\label{ccr2} \\
&a_{i_1}(k_1)\, a^{\ast i_2}(k_2) - a^{\ast i_2}(k_2)\,
a_{i_1}(k_1) = 
2\pi \delta(k_1-k_2)\, \delta_{i_1}^{i_2}\, {\bf 1}\, .  
\label{ccr3}
\eea 
The equation of motion (\ref{eqm}) and the boundary condition (\ref{bc}) follow 
directly from the representation (\ref{phi1}). The commutation relations 
(\ref{initial1},\ref{initial2}) are a consequence of (\ref{ccr1}-\ref{ccr3}) and 
the condition (\ref{compl1}). 
If the latter is violated, in addition to $\{a_i(k),\, a^{* i}(k)\, :\, k>0\}$ 
the system admits other 
degrees of freedom, which must be taken into account for reproducing 
(\ref{initial2}). These new degrees correspond to vertex bound states, 
which have been described in detail for $n=2$ in \cite{Mintchev:2004jy}. 
In order to simplify the discussion, we assume in the present paper that 
(\ref{compl1}) holds. 

In the representation (\ref{phi1},\ref{phi2}) the boundary condition 
(\ref{bc}) is captured by the wave functions $\chi$ and $\chi^*$. 
Remarkably enough, this information can be shifted to 
$a_i(k)$ and $a^{*i}(k)$, extending $\A_+$ to any 
$k\in \RR$ and considering the algebra $\A$ generated by 
$\{a_i(k),\, a^{* i}(k)\, :\, k\in \RR\}$ 
which satisfy (\ref{ccr1},\ref{ccr2}), 
\be
a_{i_1}(k_1)\, a^{\ast i_2}(k_2) - a^{\ast i_2}(k_2)\,
a_{i_1}(k_1) = 
2\pi \delta(k_1-k_2)\, \delta_{i_1}^{i_2}\, {\bf 1} +
S_{i_1}^{i_2}(k_1) 2\pi \delta(k_1+k_2)\, {\bf 1}\,  , 
\label{ccr3s}
\ee 
and the constraints   
\be
a_i(k) = S_i^j (k) a_j (-k) \, , \qquad 
a^{\ast i}(k) = a^{\ast j}(-k) S_j^i (-k)\, .    
\label{constr1}
\ee 
One can easily verify in fact that 
in terms of $\A$ the fields $\psi$ and $\psi^*$ take
the simpler form 
\be
\psi_i (t,x) = \int_{-\infty}^{+\infty} 
\frac{dk}{2\pi }
a_i (k) \e^{-\ri \omega (k)t+\ri kx}  \, , 
\label{phii1} 
\ee
\be  
\psi^{*i} (t,x) = \int_{-\infty}^{+\infty} 
\frac{dk}{2\pi }
a^{*i} (k) \e^{\ri \omega (k)t-\ri kx}  \, ,   
\label{phii2}
\ee 
where now the integration over $k$ runs over the whole line. 
The constraints (\ref{constr1}) implement in (\ref{phii1},\ref{phii2}) 
the boundary conditions (\ref{bc}) and are consistent because 
of (\ref{unit2}). Hermitian analyticity (\ref{Ha}) implies that the mapping 
\be
I\; :\; a^{\ast i}(k) \mapsto a_i (k)\, , \qquad 
I\; :\; a_i(k) \mapsto a^{\ast i} (k) \, , 
\label{involution}
\end{equation} 
generates an involution in $\A$, which is consistent with the 
second term in the right hand side of (\ref{ccr3s}). 
Because of (\ref{compl1}), the initial conditions 
(\ref{initial1},\ref{initial2}) are not influenced by this term. 
It contributes however to the time evolution and codifies 
the interaction in the vertex $V$, playing the role of a defect. 
Indeed, the algebra $\A$ defined by 
(\ref{ccr1},\ref{ccr2},\ref{ccr3s}, \ref{constr1}) is a special case of the boundary  
and the reflection-transmission algebras introduced in \cite{Liguori:1996xr} and 
\cite{Mintchev:2002zd, Mintchev:2003ue} for dealing respectively with boundaries 
and defects in quantum field theory. With our present choice of coordinates, 
the direction of both reflected and transmitted waves coincides with the 
orientation of the edges. For this 
reason both reflection and transmission amplitudes in (\ref{ccr3s}) multiply the 
$\delta(k_1+k_2)$-contribution.\footnote{This is not the case for the coordinates 
adopted for $n=2$ in \cite{Mintchev:2003ue}, where the transmission 
coefficients multiply $\delta(k_1-k_2)$ because $E_1=\RR_+$ but $E_2=\RR_-$.}

The Hamiltonian $H$ of our system is the self-adjoint extension of 
$-\prt_x^2$, defined by the pair $(A, B)$. 
The work with this somehow implicit form of $H$ is greatly simplified 
in the algebraic setting, where $H$ takes the familiar quadratic form  
\be 
H = \frac{1}{2}\int_{-\infty}^{+\infty} 
\frac{dk}{2\pi}\omega(k) a^{\ast i}(k) a_i (k) = 
\int_{0}^{+\infty} 
\frac{dk}{2\pi}\omega(k) a^{\ast i}(k) a_i (k) 
\, ,  
\label{ham}
\ee 
which is very convenient to deal with. One easily verifies for instance 
\be 
\psi_i (t,x) = \e^{\ri tH} \psi_i (0,x) \e^{-\ri tH} \,  ,  
\ee 
which confirms that (\ref{ham}) generates indeed the time evolution of $\psi$. 
It is also worth stressing that the energy is conserved, showing that 
the vertex $V$ behaves like a defect without dissipation.
\medskip 

We are interested in the fields 
$\psi$ and $\psi^*$ in the range $x\geq 0$. Nevertheless, 
eqs.(\ref{phii1},\ref{phii2}) keep a well-defined meaning for any $x\in \RR$. 
The resulting fields satisfy 
the equation of motion (\ref{eqm}), but do not obey 
the equal-time commutator (\ref{initial2}) on the whole line. 
The constraints (\ref{constr1}) imply 
\be 
\psi_i (t,x) = \int^0_{-\infty}\, dy\, {\check S}_i^j(y)\,  
\psi_j (t, y-x) \, , 
\label{pm}
\ee 
which relates the values of $\psi$ on $\RR_+$ with those on 
$\RR_-$. 
\bigskip

Summarizing, we translated so far the construction \cite{Kostrykin:1998gz} of the self-adjoint 
extension of the operator $-\prt_x^2$, relative to the boundary 
condition (\ref{bc}), in terms of the algebra $\A$. 
The advantage of the algebraic
approach is that: 

(i) it admits a straightforward extension to more general scattering 
matrices then those defined by (\ref{S1}); 

(ii) it provides a framework for introducing and studying some integrable 
interactions in the bulk of $G$. 

Let us concentrate now on point (i), dedicating to (ii) section 5 below. 
The key observation is that the algebra $\A$ is actually well-defined for 
any $S$-matrix satisfying (\ref{unit1},\ref{Ha}), which allows to 
consider much more general boundary conditions then (\ref{bc}). 
This fact is evident already for $n=1$. 
In this case the general $S$-matrix satisfying (\ref{unit1},\ref{Ha}) can be 
written in the form  
\be 
S(k) = \frac{s(k)-\ri}{s(k)+\ri} \, , \qquad s(-k) = -s(k)\, , \qquad s(k) \in \RR\, . 
\label{S3}
\ee 
Under mild technical assumptions on $s$, one can define the pseudo-differential operator 
\be 
\left [D_s f\right ](x) =  \int_{-\infty}^{+\infty} 
\frac{dk}{2\pi} s(k) {\hat f}(k) \e^{-\ri k x} \, , 
\label{psdop1}
\ee
${\hat f}$ being the Fourier transform of $f$. The function $s$ is called 
the symbol of the operator $D_s$. Setting $\e_p(x) \equiv \e^{-\ri p x}$, 
one immediately verifies that 
\be 
\left [D_s \e_p\right ](x) = s(p) \e_p(x)\, , \qquad 
\left [D_s \e_{-p}\right ](x) = s(-p) \e_{-p}(x) = -s(p) \e_{-p}(x) \, . 
\label{id1}
\ee
Using these simple identities and the representation (\ref{phi1}) with $S$ given 
by (\ref{S3}), one obtains that $\psi \equiv \psi_1$ satisfies the general boundary 
condition \cite{Mintchev:2001aq} 
\be 
\left [ \left (D_s +\ri \right )\psi  \right ](t,0) = 0 \, .  
\label{psdopbc}
\ee 
The above argument extends to star graphs with $n>1$ edges as well. One 
concludes therefore that besides (\ref{bc}), the algebra $\A$ covers also 
more general boundary conditions, formulated in terms 
of pseudo-differential operators. Computing the correlation functions of 
$\psi$ and $\psi^*$ in the next section, we apply the general framework offered 
by $\A$ without referring necessarily to (\ref{bc}). 

It is well-known that the non-relativistic field $\psi$ can be quantized with 
Fermi statistics as well, replacing the commutators (\ref{initial1},\ref{initial2}) 
with anticommutators. Obviously, for this purpose one has to perform the same 
modification in (\ref{ccr1}-\ref{ccr3s}). 

\subsection{Relativistic dispersion}

The above setting has a straightforward extension to quantum fields with 
relativistic dispersion relation. 
Let us consider for example the Hermitian scalar field $\varphi (t,x,i)$ 
defined on the graph $G$ by 
\be
\left (\prt_t^2 - \prt_x^2 + m^2 \right )\varphi (t,x,i) = 0\, , 
\qquad t\in \RR \, ,\quad  x>0 \, , \quad i=1,...,n\, , 
\label{eqmr}
\ee
and the equal-time canonical commutation relations 
\be
[\varphi (0,x_1,i_1)\, ,\, \varphi (0,x_2,i_2)] = 0  \, , 
\label{initial1r} 
\ee
\be
[(\prt_t\varphi )(0,x_1,i_1)\, ,\, \varphi (0,x_2,i_2)] =  
-i\delta_{i_1}^{i_2}\, \delta (x_1-x_2)  \, .  
\label{initial2r}
\ee
Since $\varphi$ is Hermitian, we impose on it the counterpart of the 
vertex boundary condition (\ref{bc}) for real $A$ and $B$. 
In this case one infers from (\ref{S2}) that $S$ is symmetric, i. e. 
\be 
S^t(k) = S(k) \, .  
\label{sym}
\ee
Like in the non-relativistic case, we introduce at this stage 
the algebra $\A$ with $S$ obeying  
(\ref{unit1},\ref{Ha},\ref{compl1},\ref{sym}). Then   
\be
\varphi (t,x,i) = \int_{-\infty}^{+\infty} 
\frac{dk}{2\pi \sqrt {2\omega (k)}}
\left[a^{\ast i}(k) \e^{\ri \omega (k)t-\ri kx} +
a_i (k) \e^{-\ri \omega (k)t+\ri kx}\right ] \, , 
\label{phiir}
\ee 
with  
\be 
\omega(k) = \sqrt {k^2 + m^2}\, ,   
\label{disp2}
\ee 
is the solution of (\ref{eqmr}-\ref{initial2r}) we are looking for. 
The Hamiltonian is obtained from (\ref{ham}), inserting the relativistic 
dispersion relation (\ref{disp2}). 

Summarizing, we introduced canonical quantum fields on a star 
graph $G$. The new feature, with respect to the conventional free fields on 
$\RR^s$, is the presence of a non-trivial one-particle $S$-matrix 
describing the interaction at the vertex $V$. The latter represents 
physically a point-like defect. Quantum field theory with such defects 
has been investigated \cite{Cherednik:jt}-\cite{Hallnas:2004dn} 
in various frameworks. The results of this section confirm once more 
the conclusion of \cite{Caudrelier:2004hj}, indicating the algebra $\A$ 
as a universal tool for quantization with defects. Our goal in what follows 
will be to investigate the basic features of the quantum fields constructed so far on $G$. 
Before doing that however, it is instructive to illustrate at this stage 
the vertex boundary conditions (\ref{bc}) with some examples. 

\subsection{Examples} 

The simplest choice for $A$ and $B$ in (\ref{bc}) is 
\be
A=\left(\begin{array}{cccccc}
1&-1&0& \cdots &0&0\\ 
0&1&-1& \cdots &0&0\\
\vdots&\vdots&\vdots&\cdots&\vdots&\vdots\\
0&0&0&\cdots&1&-1\\
0&0&0&\cdots&0&1
\end{array}\right)\, , \qquad B=0\, , 
\label{dir1}
\ee
which give rise to the star graph generalization 
\be 
\psi_1(t,0)=\psi_2(t,0)=\dots =\psi_n(t,0) = 0 
\label{dir2}
\ee 
of the familiar Dirichlet boundary condition. In this case $S(k) = \II_n$. 

A slightly more complicated example is 
\be
A=\left(\begin{array}{cccccc}
1&-1&0& \cdots &0&0\\ 
0&1&-1& \cdots &0&0\\
\vdots&\vdots&\vdots&\cdots&\vdots&\vdots\\
0&0&0&\cdots&1&-1\\
0&0&0&\cdots&0&-\eta 
\end{array}\right)\, , \qquad 
B=\left(\begin{array}{cccccc}
0&0&0& \cdots &0&0\\ 
0&0&0& \cdots &0&0\\
\vdots&\vdots&\vdots&\cdots&\vdots&\vdots\\
0&0&0&\cdots&0&0\\
1&1&1&\cdots&1&1 
\end{array}\right)\, , 
\label{rob1}
\ee
which leads to 
\be 
\psi_1(t,0)=\psi_2(t,0)=\dots =\psi_n(t,0)\, , \qquad 
\sum_{i=1}^n (\prt_x\psi_i) (t,0) = \eta \psi_n(t,0)\, ,  
\label{rob2}
\ee 
generalizing the Robin (mixed) boundary condition. The associated 
$S$-matrix is non-trivial, 
\be 
S(k) = \frac{1}{nk+\ri \eta}
\left(\begin{array}{ccccc}
(2-n)k-\ri \eta&2k&2k& \cdots &2k\\ 
2k&(2-n)k-\ri \eta&2k& \cdots &2k\\
\vdots&\vdots&\vdots&\cdots&\vdots\\ 
2k&2k&2k&\cdots&(2-n)k-\ri \eta  
\end{array}\right)\, .
\label{rob3}
\ee 
In particular, $\eta=0$ provides a generalization of the Neumann condition for $n\geq2$. 

Notice that the boundary conditions (\ref{dir2},\ref{rob2}) are symmetric under edge permutations. This is clearly not the case in general. A simple asymmetric example 
is defined by 
\be 
A =\frac{2 \lambda}{3} 
\left(\begin{array}{ccc}
-1&1&0\\ 
0&-1&1\\
0&0&0  
\end{array}\right)\, , \qquad 
\lambda >0\, , \qquad 
B =
\left(\begin{array}{ccc}
1&0&0\\ 
0&0&0\\
1&1&1  
\end{array}\right) \, . 
\label{ex1}
\ee 
The associated $S$-matrix reads 
\be 
S(k) = \frac{1}{3(k+\ri \lambda)} 
\left(\begin{array}{ccc}
3k-\ri \lambda&2\ri \lambda&2\ri \lambda\\ 
2\ri \lambda&-\ri \lambda&3k+2\ri \lambda\\
2\ri \lambda&3k+2\ri \lambda&-\ri \lambda  
\end{array}\right)\, , 
\label{ex2}
\ee
which shows that the boundary condition determined by (\ref{ex1}) 
is not invariant under the permutations $1\leftrightarrow 2$ and 
$1\leftrightarrow 3$. 

Being associated with critical points in the context of statistical mechanics, 
the scale invariant vertex boundary conditions are of special interest. 
The subset of such conditions of the form (\ref{bc}) has been fully described   
in \cite{Kostrykin:1998gz}. As expected, the corresponding $S$-matrices are $k$-independent. 
An example of this type, obtained by setting $\eta=0$ in (\ref{rob3}), is  
\be  
S = \frac{1}{n}
\left(\begin{array}{ccccc}
(2-n)&2&2& \cdots &2\\ 
2&(2-n)&2& \cdots &2\\
\vdots&\vdots&\vdots&\cdots&\vdots\\ 
2&2&2&\cdots&(2-n)   
\end{array}\right)\, .
\label{scaleinv}
\ee 

Let us observe finally that given a pair of matrices $(A,B)$ fixing the 
boundary condition (\ref{bc}), the pair 
$({\widetilde A}, {\widetilde B}) = (-B,A)$ defines an admissible condition 
as well. By means of (\ref{S2}), the relative scattering matrix satisfies  
\be 
{\widetilde S}(k) = S (k^{-1}) \, . 
\label{dual}
\ee 
This remarkable property \cite{Kostrykin:1998gz} shows that the mapping 
$(A, B) \mapsto (-B,A)$ is a sort of duality transformation, relating high and low energies. 
Being $k$-independent, the scale invariant $S$-matrices are obviously dual invariant. 
One should keep in mind that duality does not preserve in general 
the completeness condition (\ref{compl1}). In fact, it follows from 
(\ref{dual}) that resonant states are dual images of bound states and vice versa. 

\bigskip 

\sect{Correlation functions}

In order to extract concrete physical information from the algebra $\A$, 
one should study its representations. We focus here on two of them: 
the Fock representation $\rep$ and the Gibbs representation $\trep$ 
at inverse temperature $\beta$. Since all technical details about the construction 
and the properties of $\rep$ and $\trep$ can be found in \cite{Liguori:1996xr} 
and \cite{Mintchev:2004jy} respectively, 
we collect below only the basic formulae. Employing the commutation 
relations (\ref{ccr1},\ref{ccr2},\ref{ccr3s}), one can reduce the computation of a generic 
correlation function to correlators of the form 
\be  
\langle \prod_{k=1}^m a_{i_k}(p_{i_k}) \prod_{l=1}^n a^{\ast j_l}(q_{j_l})\rangle \, , 
\label{cf1}
\ee
which can be evaluated in turn by iteration via 
\be 
\langle \prod_{k=1}^m a_{i_k}(p_{i_k}) \prod_{l=1}^n a^{\ast j_l}(q_{j_l})
\rangle = 
\delta_{mn}\, \sum_{k=1}^m \langle a_{i_1}(p_{i_1})a^{\ast j_k}(q_{j_k})\rangle  
\, \langle \prod_{k=2}^m a_{i_k}(p_{i_k}) 
\prod_{\stackrel{l=1}{l\not=k} }^n a^{\ast j_k}(q_{j_k}) \rangle \, .   
\label{cf2}
\ee
{} For this purpose one needs only the two-point functions. In $\rep$ on has 
\be
\langle a_i(p)a^{\ast j}(q)\rangle  = 
2\pi \left [\delta_i^j\, \delta (p-q) 
+ S_i^j(p)\, \delta (p+q)  \right ]\,  , \qquad 
\langle a^{\ast i}(p )a_j(q)\rangle = 0\, .   
\label{fock1}
\ee
In $\trep$ one finds \cite{Mintchev:2004jy} instead 
\bea 
\langle a_i(p)a^{\ast j}(q)\rangle_\beta  =
\frac{1}{ 1 \pm \e^{-\beta [\omega(p)-\mu]}} \, 
2\pi \left[\delta_i^j \delta (p-q) 
+ S_i^j(p)\delta (p+q)  \right ]\,  ,
\label{gibbs1}\\
\langle a^{\ast i}(p)a_j(q)\rangle_\beta  =
\frac{\e^{-\beta [\omega(p)-\mu]}}{ 1 \pm \e^{-\beta [\omega(p)-\mu]}}\,  
2\pi \left[\delta^i_j \delta (p-q) 
+ S^i_j(-p)\delta (p+q)  \right ]\,  ,
\label{gibbs2}
\eea 
where $\pm$ stands for Fermi/Bose statistics and 
$\mu \in \RR$ is the chemical potential associated with the number operator 
\be
N = \frac{1}{2}\int_{-\infty}^{+\infty} 
\frac{dk}{2\pi} a^{\ast i}(k) a_i (k) \, . 
\label{num}
\ee 
A characteristic property of any thermal representation is the 
Kubo-Martin-Schwinger (KMS) condition. In our context it states that 
\be
\langle \left [\alpha_s a_{i}(p)\right ]a^{\ast j}(q)\rangle_\beta  =
\langle a^{\ast j}(q)\left [\alpha_{s+\ri \beta} a_{i}(p)\right ]\rangle_\beta  \, , 
\label{KMS}
\ee 
$\alpha_s$ being the automorphism 
\be
\alpha_s \, a^{\ast i}(k) = 
\e^{\ri sK} a^{\ast i}(k) \e^{-\ri sK} \, , \qquad
\alpha_s \, a_i (k) = \e^{\ri sK} a_i (k) \e^{-\ri sK}\, ,   
\qquad K \equiv H - \mu N \, . 
\ee
One easily verifies that (\ref{gibbs1},\ref{gibbs2}) satisfy (\ref{KMS}). 

It is straightforward now to derive the two-point correlators of the fields 
$\psi$, $\psi^*$ and $\varphi$ introduced in the previous section. 
In the Fock representation $\rep$ one gets 
\bea
\langle \psi^{*i_1}(t_1,x_1) \psi_{i_2} (t_2,x_2)\rangle = 
\qquad \qquad \qquad \quad 
\nonumber \\
\langle \psi_{i_1} (t_1,x_1) \psi_{i_2} (t_2,x_2)\rangle = 
\langle \psi^{*i_1}(t_1,x_1) \psi^{*i_2}(t_2,x_2)\rangle = 0 \, ,
\label{psi22} 
\eea 
\be
\langle \psi_{i_1}(t_1,x_1) \psi^{*i_2}(t_2,x_2)\rangle 
= \int_{-\infty}^{+\infty}\frac{dk}{2\pi} \e^{-\ri \omega (k) t_{12}} 
\left [\e^{\ri k x_{12}} \delta_{i_1}^{i_2} + \e^{\ri k{\widetilde x}_{12}} 
S_{i_1}^{i_2}(k)\right ] \, , 
\label{psi21}
\ee
with 
\be 
x_{12}\equiv x_1-x_2\, ,\quad {\widetilde x}_{12} \equiv x_1+x_2
\ee 
and the dispersion relation (\ref{disp1}). 
In the relativistic case (\ref{disp2}) one obtains 
\be
\langle \varphi(t_1,x_1,i_1) \varphi(t_2,x_2,i_2)\rangle 
= \int_{-\infty}^{+\infty}\frac{dk}{4\pi \omega(k)} \e^{-\ri \omega (k) t_{12}} 
\left [\e^{\ri k x_{12}} \delta_{i_1}^{i_2} + \e^{\ri k{\widetilde x}_{12}} 
S_{i_1}^{i_2}(k)\right ] \, .  
\label{phi21}
\ee 
We see that the fields in the different edges are 
not independent, but interact through the scattering matrix $S$ 
taking into account the boundary conditions in the vertex of the graph. 

In the Gibbs representation $\trep$ one finds\footnote{In the bosonic case 
we assume that $\mu < 0$.} 
\be
\langle \psi_{i_1}(t_1,x_1) \psi_{i_2} (t_2,x_2)\rangle_\beta =
\langle \psi^{*i_1}(t_1,x_1) \psi^{*i_2}(t_2,x_2)\rangle_\beta = 0 \, ,  
\label{tpsi23} 
\ee
\bea
\langle \psi_{i_1}(t_1,x_1) \psi^{*i_2}(t_2,x_2)\rangle_\beta = 
\qquad \qquad \qquad 
\nonumber \\ 
\int_{-\infty}^{+\infty}\frac{dk}{2\pi} \frac{1}{ 1 \pm \e^{-\beta [\omega(k)-\mu]}} 
\e^{-\ri \omega (k) t_{12}} 
\left [\e^{\ri k x_{12}} \delta_{i_1}^{i_2} + \e^{\ri k{\widetilde x}_{12}} 
S_{i_1}^{i_2}(k)\right ] \, , 
\label{tpsi21}
\eea
\bea
\langle \psi^{*i_1}(t_1,x_1) \psi_{i_2} (t_2,x_2)\rangle_\beta = 
\qquad \qquad \qquad 
\nonumber \\ 
\int_{-\infty}^{+\infty}\frac{dk}{2\pi} 
\frac{\e^{-\beta [\omega(k)-\mu]}}{ 1 \pm \e^{-\beta [\omega(k)-\mu]}} 
\e^{\ri \omega (k) t_{12}} 
\left [\e^{\ri k x_{12}} \delta^{i_1}_{i_2} + \e^{\ri k{\widetilde x}_{12}} 
S^{i_1}_{i_2}(k)\right ] \, , 
\label{tpsi22}
\eea 
\bea
\langle \varphi(t_1,x_1,i_1) \varphi(t_2,x_2,i_2)\rangle_\beta = 
\qquad \qquad \qquad \qquad \quad 
\nonumber \\
\int_{-\infty}^{+\infty}\frac{dk}{4\pi \omega(k)} 
\frac{\e^{-\beta [\omega(k)-\mu]+i\omega (k)t_{12}} +
\e^{-i\omega (k)t_{12}}} { 1-\e^{-\beta [\omega(k)-\mu]}}  
\left [\e^{\ri k x_{12}} \delta_{i_1}^{i_2} + \e^{\ri k{\widetilde
x}_{12}}  S_{i_1}^{i_2}(k)\right ] \, .  
\label{tphi21}
\eea 

With the above background we are ready to compute some quantum field 
theory observables like the particle, current and energy densities on a star graph. 

\bigskip 

\sect{Observables} 

The basic local observables associated with the field $\psi$ are the particle density 
\begin{equation}
\varrho(t,x,i)=\left[\psi^{*i}\psi_i\right](t,x)\, ,   
\label{dens1}
\end{equation} 
the current density 
\begin{equation}
j(t,x,i)= \frac{\ri }{2m}\left [ \psi^{*i}(\partial_x\psi_i) - (\partial_x\psi^{*i})\psi_i \right ](t,x) \, , 
\label{currdens1}
\end{equation}
and the energy density 
\begin{equation}
\theta (t,x,i) = -\frac{1}{4m}\left[
\psi^{*i}\left (\partial_x^2 \psi_i \right )+
\left (\partial_x^2 \psi^{*i} \right )\psi_i \right](t,x) \, , 
\label{endens1} 
\end{equation}
without summation over $i$ in the right hand side. Conservation 
and Kirchoff's laws 
\be
\prt_t \, \varrho(t,x,i) = \prt_x \, j(t,x,i) \, , \qquad 
\sum_{i=1}^{n}j(t,0,i)=0 \, , 
\ee 
\label{laws1}
{}follow from (\ref{eqm}) and (\ref{bc}) respectively. 

It is straightforward to compute the expectation values of (\ref{dens1}-\ref{endens1}) in the 
Gibbs representation $\trep$ defined in the previous section. 
Applying the conventional point-splitting procedure, one gets from (\ref{tpsi22})  
\bea
\langle \varrho(t,x,i) \rangle_{\beta}= 
\lim_{t_1\to t_2 =\, t \atop x_1\to x_2 =\, x} 
\langle \psi^{*i_1}(t_1,x_1) \psi_{i_2} (t_2,x_2)\rangle_\beta = 
\qquad \qquad \nonumber \\
\int_{-\infty}^{+\infty}\frac{dk}{2\pi}
\;\frac{e^{-\beta[\omega(k)-\mu]}}
{1 \pm e^{-\beta[\omega(k)-\mu]}}[1+S^i_i(k)e^{2\ri kx}]\, , \qquad 
\omega(k) = \frac{k^2}{2m}\, . 
\qquad  
\label{tdens1}
\eea
Analogously, $\langle j(t,x,i) \rangle_{\beta}= 0$ and 
\begin{equation}
\langle \theta (t,x,i) \rangle_{\beta}= 
\int_{-\infty}^{+\infty}\frac{dk}{2\pi}\, \omega(k) 
\;\frac{e^{-\beta[\omega(k)-\mu]}}
{1 \pm e^{-\beta[\omega(k)-\mu]}}[1+S^i_i(k)e^{2\ri kx}]\, .   
\label{tdens2}
\end{equation} 
Egs. (\ref{tdens1},\ref{tdens2}) provide convenient integral representations of 
the particle and energy densities in the Gibbs state. Since the integrands 
involve only diagonal elements of the $S$-matrix, 
the expectation values (\ref{tdens1},\ref{tdens2}) are real because 
of Hermitian analyticity (\ref{Ha}). In agreement with the invariance of the Gibbs state 
under time translations, (\ref{tdens1},\ref{tdens2}) 
are $t$-independent and vanish in the limit of zero 
temperature ($\beta \to \infty $), which shows that these densities have a 
purely thermal origin. 

Let us concentrate on the energy density (\ref{tdens2}). 
It is instructive to separate the universal
contribution 
\be 
\varepsilon_\pm (\beta) = 
\int_{-\infty}^{+\infty}\frac{dk}{2\pi}\, \omega(k) 
\;\frac{e^{-\beta[\omega(k)-\mu]}}{1 \pm e^{-\beta[\omega(k)-\mu]}} \, , 
\label{sb1}
\ee 
{}from the vertex dependent part 
\be
\E_\pm (x,i,\beta) = \int_{-\infty}^{+\infty}\frac{dk}{2\pi}\, \omega(k) 
\;\frac{e^{-\beta[\omega(k)-\mu]}}{1 \pm e^{-\beta[\omega(k)-\mu]}}\, S^i_i(k)e^{2\ri kx}\, . 
\label{vc1}
\ee 
Unitarity (\ref{unit1}) of $S$ implies the estimate 
 \begin{equation}
| \E_\pm (x,i,\beta)|\leq \varepsilon_\pm (\beta) \, , 
\label{estim2} 
 \end{equation} 
leading in turn to positivity 
\begin{equation}
\langle \theta (t,x,i) \rangle_{\beta} = \varepsilon_\pm (\beta) +  
\E_\pm (x,i,\beta)  \geq 0 
\label{estim3}
\end{equation}
of the energy density in the Gibbs state. The integration over $k$ in (\ref{sb1}) gives 
\be 
\varepsilon_\pm (\beta) = \mp \frac{1}{2m\sqrt {2\pi}} 
{\rm Li}_{\frac{3}{2}}\left (\mp \e^{\beta \mu}\right ) \left (\frac{m}{\beta}\right )^{\frac{3}{2}} \ , 
\label{sb2}
\ee 
where ${\rm Li}_s$ is the polylogarithm function. Eq. (\ref{sb2}) is the analog of the  
Stefan-Boltzmann (S-B) law for the non-relativistic dispersion relation (\ref{disp1}). With 
$\mu =0$ it simplifies further  
\be 
\varepsilon_- (\beta) =\frac{1}{2m\sqrt{2\pi}}\, \zeta \left (\frac{3}{2}\right )
\left (\frac{m}{\beta}\right )^{\frac{3}{2}} \, , 
\quad 
\varepsilon_+ (\beta) =\frac{\sqrt{2}-1}{2m\sqrt{2\pi}}\, \zeta \left (\frac{3}{2}\right )
\left (\frac{m}{\beta}\right )^{\frac{3}{2}} \, , 
\label{sb3}
\ee
$\zeta$ being Riemann's zeta function. 

According to the previous discussion, the vertex dependent 
contribution (\ref{vc1}) represents a correction to the 
S-B law generated by the defect at $x=0$.  
As expected, this term is both 
$x$ and $\beta$-dependent and involves the $S$-matrix. The explicit 
integration in $k$ is usually a hard problem, but the 
integral representation (\ref{vc1}) is very convenient for the numerical 
evaluation of $\E_\pm (x,i,\beta)$. 

Let us consider now the case with relativistic dispersion (\ref{disp2}), focusing on 
the scalar field (\ref{eqmr}-\ref{initial2r}). The energy density in this case reads 
\be
\theta (t,x,i)  =  \frac{1}{2}\left \{ (\partial_t\varphi)^2 -  
\frac{1}{2}\left [ \varphi(\partial^2_x\varphi) + (\partial^2_x\varphi) \varphi \right ] + 
m^2 \varphi^2 \right \}(t,x,i)\, . 
\label{edens2}
\ee 
One can compute the expectation value $\langle \theta (t,x,i) \rangle_{\beta}$ via 
point-splitting from the two-point function (\ref{tphi21}). Differently from the non-relativistic case, 
it turns out that $\langle \theta (t,x,i) \rangle_{\beta}$ diverges. This is not surprising because 
the same problem is encountered already in the standard Casimir effect. 
Like in that case, one can solve it by subtracting from $\langle \theta (t,x,i) \rangle_{\beta}$ 
the expectation value $\langle \theta (t,x) \rangle_{\infty}^{\rm line}$ of the energy density 
$\theta (t,x)$ on the whole line $\RR$ and at zero temperature. The result is  
\be 
\langle \theta (t,x,i) \rangle_{\beta} - 
\langle \theta (t,x) \rangle_{\infty}^{\rm line} = 
\varepsilon_{{}_{\rm S-B}} (\beta) + \E_{{}_{\rm C}} (x,i) + \E(x,i,\beta ) \, , 
\label{endens3}
\ee 
where 
\be 
\varepsilon_{{}_{\rm S-B}} (\beta) = 
\int_{-\infty}^{+\infty}\frac{dk}{2\pi}\, \omega(k) 
\;\frac{e^{-\beta[\omega(k)-\mu]}}{1 - e^{-\beta[\omega(k)-\mu]}} \, , 
\qquad \omega(k) = \sqrt{k^2+m^2}\, , 
\label{sb4}
\ee 
is the Stefan-Boltzmann contribution, 
\be
\E_{{}_{\rm C}} (x,i) = \frac{1}{2} \int_{-\infty}^{+\infty}\frac{dk}{2\pi}\, \omega(k)\, S^i_i(k)e^{2\ri kx}   
\label{vc2}
\ee 
is the Casimir energy density associated with the vertex interaction at zero temperature and 
\be
\E (x,i,\beta) = \int_{-\infty}^{+\infty}\frac{dk}{2\pi}\, \omega(k) 
\;\frac{e^{-\beta[\omega(k)-\mu]}}{1- e^{-\beta[\omega(k)-\mu]}}\, S^i_i(k)e^{2\ri kx}\,  
\label{vc3}
\ee 
is the finite temperature correction to the latter. 

It is now easy to analyze separately the three contributions (\ref{sb4}-\ref{vc3}). 
To the end of this section we set for simplicity $m=\mu =0$. Then 
\be 
\varepsilon_{{}_{\rm S-B}} (\beta) =  \frac{\pi}{6 \beta^2}\, , 
\ee 
which is the familiar S-B thermal energy density for a massless 
Hermitian scalar field in 1+1 space-time dimensions. 

Eq. (\ref{vc2}) is a useful integral representation of the Casimir energy density 
in terms of the scattering matrix $S(k)$. This representation extends the result of 
\cite{Fulling:2005js} to all boundary conditions of the type (\ref{bc}). 
In the scale invariant case for instance, 
$S(k)$ is a constant matrix $S$ leading to 
\be 
\E_{{}_{\rm C}} (x,i) = -\frac{S_i^i}{8\pi x^2} \, . 
\label{vc4}
\ee 
The corresponding finite temperature correction (\ref{vc3}) reads  
\be 
\E (x,i,\beta) =  \frac {\pi S_i^i}{2 \beta^2\, {\sinh} \left (2\pi\frac{x}{\beta} \right )} 
-\frac{S_i^i}{8\pi x^2} \, . 
\label{vc5}
\ee

We illustrate the case of a $k$-dependent $S$-matrix on the example 
(\ref{ex1}) of asymmetric boundary conditions in the vertex $V$. 
The $k$-integration in (\ref{vc2}) with the diagonal elements of 
(\ref{ex2}) can be performed explicitly and one finds 
\be 
\E_{\rm C} (x,1) = -\frac{1}{8\pi x^2}
+ \frac{\lambda }{3\pi x} +
\frac{2\lambda^2 \e^{2\lambda x}}{3\pi}\, {\rm Ei}(-2\lambda x)\,  ,  
\label{ex3}
\ee 
\be
\E_{\rm C} (x,i) = 
\frac{\lambda }{12\pi x} +
\frac{\lambda^2 \e^{2\lambda x}}{6\pi}\, {\rm Ei}(-2\lambda x)\,  ,  
\qquad i=2,3\, ,
\label{ex4}
\ee 
${\rm Ei}$ being the exponential-integral function. We have plotted 
these functions for $\lambda = 1$ in Fig. 2, which shows that 
the Casimir energy in the edge $E_1$ behaves quite differently from that in 
the edges $E_{2,3}$. In fact, close to the vertex along $E_1$ one has an 
attractive force, whereas the force along $E_{2,3}$ is always repulsive. 

\begin{figure}[h]
\begin{center}
\includegraphics[scale=0.6]{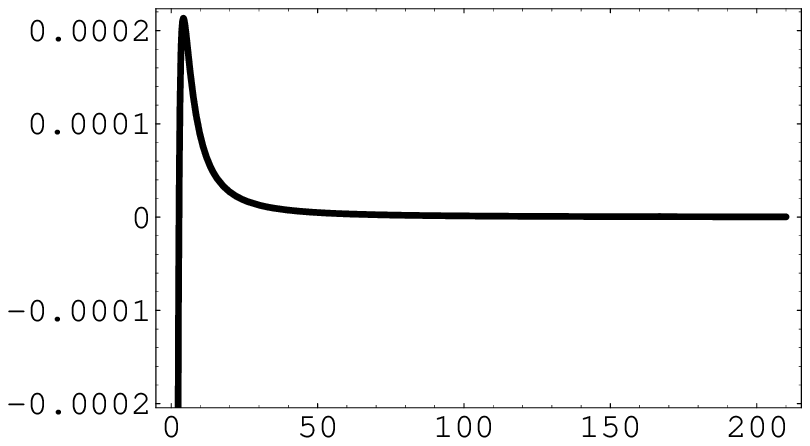}
\hskip 1.5 truecm 
\includegraphics[scale=0.6]{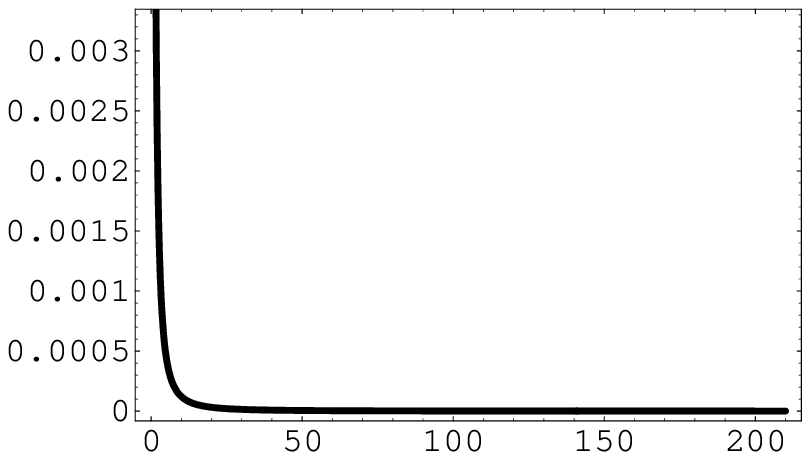}
\end{center}
\centerline{Figure 2: Plots of $\E_{\rm C} (x,1)$ (left) and 
$\E_{\rm C} (x,2)=\E_{\rm C} (x,3)$ (right) at $\lambda =1$.} 
\end{figure}

Using the results of \cite{Mintchev:2004jy}, the thermal correction (\ref{vc3}) in this case equals 
\bea 
\E (x,1,\beta) =  \qquad \qquad \qquad  \qquad \qquad \qquad 
\nonumber \\
\frac {\pi }{2 \beta^2\, {\sinh} \left (2\pi\frac{x}{\beta} \right )}  -\frac{1}{8\pi x^2} 
- \frac{\lambda}{3\pi x} - \frac{2\lambda^2 \e^{2x\lambda }}{3 \pi}\, {\rm Ei}(-2\lambda x) + 
\frac{4}{\beta^2} F\left (\frac{\beta \lambda}{2\pi}, \e^{-4 \pi \frac{x}{\beta}} \right )\, ,  
\label{ex5} 
\eea
\be 
\E (x,i,\beta)  = 
- \frac{\lambda}{12\pi x} - \frac{\lambda^2 \e^{2x\lambda }}{6 \pi}\, {\rm Ei}(-2\lambda x) + 
\frac{1}{\beta^2} F\left (\frac{\beta \lambda}{2\pi}, \e^{-4 \pi \frac{x}{\beta}} \right )\, , 
\qquad  i=2,3\, ,   
\label{ex6} 
\ee
with
\be 
F(\sigma, \tau) = \frac{2 \pi \sigma \tau}{3 (1 + \sigma) }\, \,  
{}_2{\rm F}_1\left [2, 1+\sigma;  2 +\sigma; \tau \right ]\, , 
\label{hyp}
\ee
where ${}_2{\rm F}_1$ is the hypergeometric function. Fig. 3 and Fig. 4 display the behavior 
of $\E (x,i,\beta)$ at fixed $\beta$ and $x$ respectively.  

\begin{figure}[h]
\begin{center}
\includegraphics[scale=0.6]{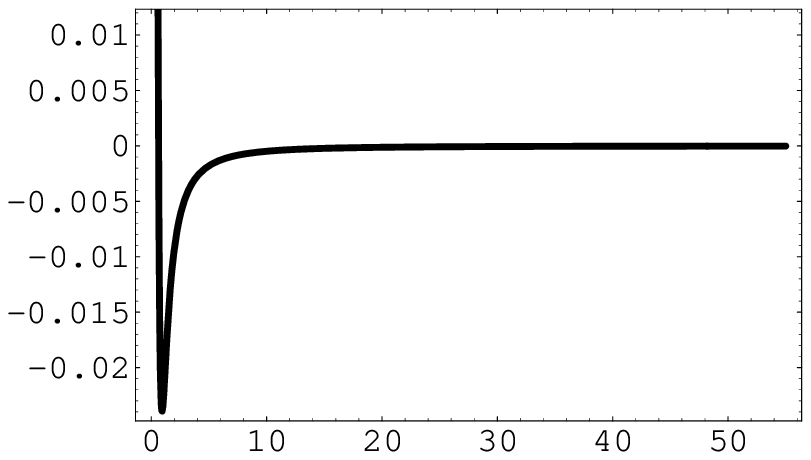}
\hskip 1.5 truecm 
\includegraphics[scale=0.6]{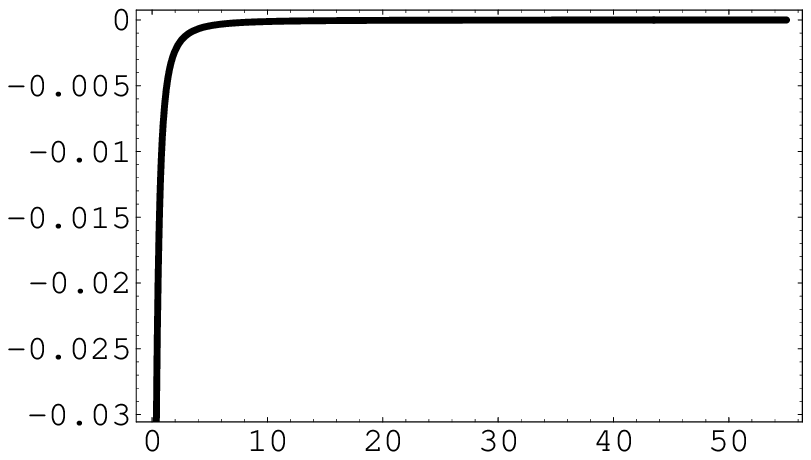}
\end{center}
\centerline{Figure 3: Plots of $\E (x,1,1)$ (left) and $\E(x,2,1)=\E(x,3,1)$ 
(right) at $\lambda =1$.}  
\end{figure} 

\begin{figure}[h]
\begin{center}
\includegraphics[scale=0.6]{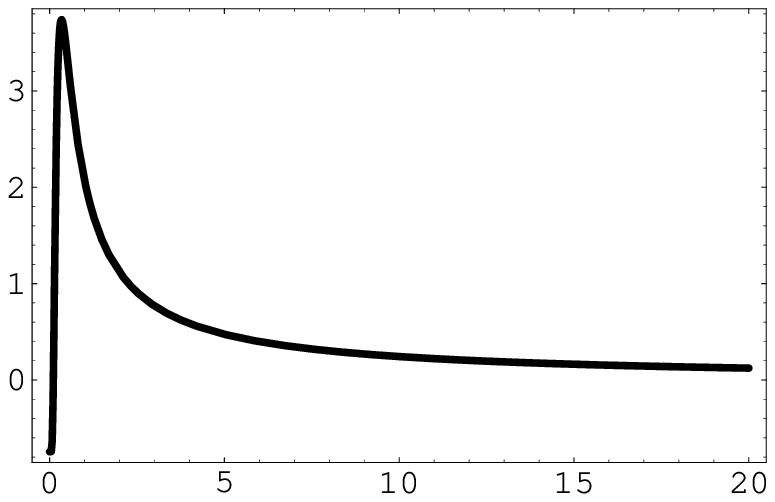}
\hskip 1.5 truecm 
\includegraphics[scale=0.6]{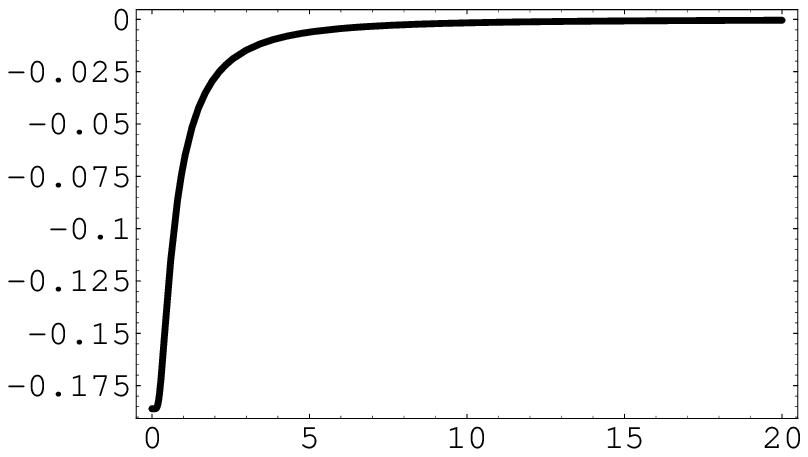}
\end{center}
\centerline{Figure 4: Plots of $\E (0.1,1,\beta )$ (left) and $\E(0.1,2,\beta )=\E(0.1,3,\beta)$ 
(right) at $\lambda =1$.}  
\end{figure}

Summarizing, we derived in this section the expectation values of some local observables in 
the Gibbs state, paying attention to the contribution of the vertex interaction captured by  
the boundary conditions (\ref{bc}). In what follows we will address the problem of bulk interactions.

\bigskip 
\bigskip

\sect{Non-trivial bulk interactions} 

The dynamics of the classical Non-Linear Schr\"odinger (NLS) model 
on a star graph $G$ is described by the equation of motion\footnote{In order 
to simplify the notation, in this section we fix the units in such a way that $m=\frac{1}{2}$.}  
\be
\left (\ri \prt_t + \prt_x^2\right )\psi_i (t,x) = 
2 g\, |\psi_i (t,x)|^2\, \psi_i (t,x)\, ,  
\label{eqmnls}
\ee 
$g\in \RR$ being the coupling constant. 
In order to avoid the presence of bound states in the bulk, 
we take below $g>0$.
Our main goal now will be to quantize (\ref{eqmnls}) at zero temperature, 
imposing the initial conditions (\ref{initial1},\ref{initial2}) and 
the boundary condition (\ref{bc}). For this purpose one must:    
\medskip

(i) construct a Hilbert space $\cal H$ describing the states of
the system;

\medskip
(ii) define on an appropriate dense domain $\D \subset \cal H$
the operator valued distributions $\psi_i(t,x)$ and $\psi^{*i}(t,x)$, satisfying 
(\ref{initial1},\ref{initial2},\ref{bc}) and 
the equation of motion (\ref{eqmnls}) with suitably defined operator 
product in the right hand side; 

\medskip
(iii) exhibit a vacuum state $\Omega \in \D$, 
which is cyclic with respect to the fields $\psi^{*i}(t,x)$.
\medskip 

This program has been carried out for $n=1$ and $n=2$ in 
\cite{Gattobigio:1998si} and \cite{Caudrelier:2004xy} respectively. 
Since the approach used there has a direct 
generalization to a star graph $G$ with arbitrary $n$, we will sketch only 
the main points of the construction. The first step is to generalize the algebra 
$\A$ in order to incorporate both non-trivial bulk and vertex interactions. Following 
\cite{Faddeev:gh}-\cite{Faddeev:zy}, we introduce the two-body bulk scattering 
matrix\footnote{We adopt the conventional tensor notation.} 
\be 
R_{12} (k) = \frac{k-\ri g}{k+\ri g}\,  \II_n \otimes \II_n  \, , 
\label{bulks}
\ee 
which satisfies unitarity 
\be
R_{12} (k)\, R_{21}(-k) = \II_n\otimes \II_n \, ,
\label{unit3}
\ee
Hermitian analyticity
\be
R^*_{12}(k) = R_{21}(-k) \, ,
\ee
and the quantum Yang-Baxter equation
\be
R_{12}(k_1-k_2) R_{13}(k_1-k_3) R_{23} (k_2-k_3)
= R_{23}(k_2-k_3) R_{13}(k_1-k_3) R_{12}(k_1-k_2)  \, . 
\label{qyb}
\ee
In the spirit of factorized scattering \cite{Faddeev:gh}-\cite{Faddeev:zy} 
we replace (\ref{ccr1},\ref{ccr2}) by 
\bea
a_{i_1}(k_1) \, \, a_{i_2 }(k_2) \, \; - \; \,
       R_{i_2 i_1 }^{j_2 j_1 }
       (k_2 -k_1)\,\, a_{j_2 }(k_2)\, a_{j_1 }(k_1) = 0
       \, , \qquad \quad \label{aa}\\
a^{\ast i_1} (k_1)\, a^{\ast i_2 } (k_2) -
       a^{\ast j_2 } (k_2)\, a^{\ast j_1 } (k_1)\,
       R_{j_2 j_1 }^{i_2 i_1 }(k_2 -k_1) = 0
       \, . \qquad \quad \label{a*a*}
\eea  
encoding in this way the bulk interaction in the exchange relations. 
On the other hand, in order to describe 
the vertex interaction we introduce \cite{Liguori:1996xr} 
the so called {\it boundary} generators 
$b_i^j(k)$, which appear in the right hand side of 
\bea
a_{i_1 }(k_1)\, a^{\ast i_2 } (k_2) \; - \;
         a^{\ast j_2 }(k_2)\,
         R_{i_1 j_2 }^{j_1 i_2}(k_1 - k_2)\,
         a_{j_1 }(k_1) & = &  \nonumber\\
         2\pi\, \delta (k_1 - k_2)\, \delta_{i_1}^{i_2} \, {\bf 1} + 
         2\pi\, \delta (k_1 + k_2)\,  b_{i_1 }^{i_2 }(k_1)
         \, ,
\label{aa*}
\eea 
being the counterpart of (\ref{ccr3}). Due to the 
particular form (\ref{bulks}) of $R_{12}$, 
the boundary generators commute 
\be 
[b_{i_1}^{j_1}(k_{1})\, ,\, b_{i_2}^{j_2}(k_{2})]=0\, , 
\label{bcomm}
\ee 
but have non-trivial exchange relations with $a_i(k)$ and $a^{*i}(k)$: 
\be
a_{i_1}(k_1)\, b_{i_2}^{j_2}(k_2) =
R_{i_2 i_1}^{k_2 k_1}(k_2 - k_1)\,
b_{k_2}^{l_2}(k_2)\,
R_{k_1 l_2}^{l_1 j_2}(k_1 + k_2)\,
a_{l_1}(k_1)
\, , \qquad 
\label{ab}
\ee
\be
b_{i_1}^{j_1}(k_1)\, a^{\ast i_2}(k_2) =
a^{\ast l_2}(k_2)\,
R_{i_1 l_2}^{l_1 k_2}(k_1 - k_2)\,
b_{l_1}^{k_1}(k_1)\,
R_{k_2 k_1}^{i_2 j_1}(k_2  + k_1)\,
\, . \qquad 
\label{ba*}
\ee 

The next step is to focus on the Fock representation $\rep$ of the above algebra $\A$. 
According to \cite{Liguori:1996xr}, $\rep$ is fixed (up to unitary equivalence) by the 
condensate $\langle \Omega\, ,\, b_i^j(k) \Omega \rangle$, 
where $\Omega$ is the vacuum state and $\langle \cdot\, ,\, \cdot \rangle $ 
the scalar product in $\rep$. 
In order to satisfy the boundary condition (\ref{bc}), we require that 
\be 
\langle \Omega\, ,\, b_i^j(k) \Omega \rangle = S_i^j(k)    
\label{fix}
\ee 
with $S(k)$ given by (\ref{S1}). 

The final step is to define the basic structures, described in 
(i)-(iii) above, in terms of $\rep$ as follows:

\begin{itemize}

\item {} $\Hil$ and $\Omega$ are the Hilbert space and the vacuum state of
$\rep$.

\item {} The quantum field $\psi_i(t,x)$ admits the series representation 
\be
\psi_i (t, x) =  \sum_{m=0}^\infty (-g)^m\, \psi^{(m)}_i (t,x) \, ,
\label{psiseries}
\ee 
where
\bea
\psi^{(m)}_i (t,x) =
\int_{-\infty}^\infty \prod_{k=1\atop l=0}^m
\frac{dp_k}{2\pi}\frac{dq_l}{2\pi}\,
a^{* i}(p_1)\ldots a^{* i}(p_m)a_i(q_m)\ldots
a_i(q_0) \cdot \nonumber \\
\frac{e^{\ri \sum\limits_{l=0}^m(q_l x-q^2_l t)-\ri \sum\limits_{k=1}^m(p_k x-p_k^2 t)}}
{\prod\limits_{k=1}^m(p_k-q_{k-1} - \ri \varepsilon)
(p_k-q_k - \ri \varepsilon)} \, ,
\qquad \qquad \qquad
\label{psin}
\eea
{\it without} summation over the upper and lower index $i$ in the right hand side.  

\item {} The domain $\D$ is the finite particle subspace of $\rep$,
which is well-known to be dense in $\Hil$.

\end{itemize}

\noindent It is worth mentioning that the series (\ref{psiseries}) is actually
a finite sum when $\psi_i(t,x)$ is acting on $\D$. The coupling 
constant $g$ appears explicitly in (\ref{psiseries}) and implicitly in 
$a_i(k)$ and $a^{*i}(k)$, which depend on $g$ through the bulk 
scattering matrix (\ref{bulks}). Eq. (\ref{psin}) defines a sort of 
non-linear Fourier transform of the type used in \cite{Ros, Fokas:1995rj} to solve 
the classical NLS model. In oder to give meaning 
of the equation of motion (\ref{eqmnls}) at the level of quantum fields 
we perform the substitution
\be 
|\psi_i (t,x)|^2 \psi_i (t,x)\, \mapsto \; : \psi_i \psi^{*i} \psi_i : (t,x) \, , 
\label{subst}
\ee
(without summation over $i$) 
introducing the normal ordered product $: \cdots :$ as follows \cite{Gattobigio:1998si}: 
all creation operators in such a product stand as usual to the left of 
all annihilation operators. In view of eqs. (\ref{aa},\ref{a*a*}), 
in our case one must further specify the ordering of creators and annihilators 
themselves. We define $: \cdots :$ to preserve the original order 
of the creators. The original order of the annihilators is 
preserved if both belong to the same $\psi $ or $\psi^*$ 
and inverted otherwise. With this convention one can show \cite{Gattobigio:1998si} that 
\be
(i\prt_t + \prt_x^2 )\langle \chi_1 \, ,\, \psi_i (t,x) \chi_2 \rangle
= 2g\, \langle \chi_1 \, ,\, : \psi_i \psi^{*i} \psi_i : (t,x) \chi_2 \rangle \, ,
\quad x>0,\, \, i=1,...,n , 
\label{qeqm}
\ee
holds for any $\chi_{1,2} \in \D$. The boundary condition (\ref{bc}) 
is also satisfied in mean value on $\D$, namely 
\be 
\langle \chi_1\, ,\, \left [
A_i^j \psi_j (t,0) + B_i^j (\prt_x\psi_j ) (t,0) \right ] \chi_2 \rangle = 0 \, , \quad \forall\,  t\in \RR\, ,  
\quad \chi_{1,2} \in \D\, .  
\label{bcnls} 
\ee 
{}Finally, by means of (\ref{aa}-\ref{ba*}) one can verify the 
commutation relations (\ref{initial1},\ref{initial2}). 
Since $\psi $ and $\psi^*$ are unbounded operators,
the subtle points in proving the above statements are essentially
domain problems. They are faced \cite{Gattobigio:1998si} using the absence of poles 
in the analytic extension of the boundary and the bulk scattering matrices 
(\ref{S1}) and (\ref{bulks}) in the upper complex $k$-plane. The lack 
of both bulk ($g>0$) and vertex (see eq.(\ref{compl1})) 
bound states is therefore crucial. 

The representation defined by (\ref{psiseries},\ref{psin}) allows to compute 
all correlation functions. From the general structure of the 
solution one infers that the nonvanishing correlators involve
equal number of $\psi_i $ and $\psi^{*i}$. Moreover, for deriving 
the exact $2n$-point function one does not need all terms in the 
expansion (\ref{psiseries}), but at most the $(n-1)$-th order contribution. 
In fact, 
\be 
\langle \psi_{i_1}(t_1,x_1)\psi^{*i_2}(t_2,x_2) \rangle = 
\langle \psi_{i_1}^{(0)}(t_1,x_1)\psi^{*i_2(0)}(t_2,x_2) \rangle \, , 
\ee
\bea
\langle \psi_{i_1}(t_1,x_1)\psi_{i_2}(t_2,x_2)
\psi^{*i_3}(t_3,x_3)\psi^{*i_4}(t_4,x_4) \rangle = \quad 
\nonumber \\
\langle  \psi_{i_1}^{(0)}(t_1,x_1)\psi_{i_2}^{(0)}(t_2,x_2)
\psi^{*i_3(0)}(t_3,x_3)\psi^{*i_4(0)}(t_4,x_4) \rangle + 
\nonumber \\
g^2 \langle \psi_{i_1}^{(0)}(t_1,x_1)\psi_{i_2}^{(1)}(t_2,x_2)
\psi^{*i_3(1)}(t_3,x_3)\psi^{*i_4(0)}(t_4,x_4) \rangle 
\, ,  
\eea 
and so on. Since the vacuum expectation value of any number of 
$\{a_i(k), \, a^{*i}(k),\, b_i^j(k)\}$ is known explicitly, 
one derives in this way an integral representation for the $2n$-point function 
involving $n$ momentum integrations. The correlation functions involving 
the particle density (\ref{dens1}) and the current (\ref{currdens1}) can 
be computed analogously. It will be interesting to extend this framework 
to finite temperature. 

The application of the above solution to the study of non-linear effects in the 
wave propagation on star graph quantum wires deserves further investigations.

\sect{Outlook and conclusions}  

We described in the present paper a construction of quantum fields propagating 
on a star graph $G$ with any number of edges. Our construction is 
based on a specific deformation $\A$ of the algebra of canonical 
commutation relations, which takes into account both vertex 
and bulk interactions. We consider the case without dissipation, 
leading to unitary theories. Two applications 
of the general framework have been discussed. 
The first one concerns the derivation of the expectation values 
of some observables at finite temperature. Focusing mainly 
on the energy density, we described the Casimir effect on a star 
graph $G$ and derived the correction to the Stefan-Boltzmann law 
due to the interaction in the vertex $V\in G$. 
As a second application we solved the quantum non-linear 
Schr\"odinger equation on $G$. This example illustrates our approach 
at work when integrable interactions are present in the bulk. 

The results of the paper can be generalized in several directions. First of 
all one can increase  the dimensions, setting $E_i = \RR^s$ and 
$V=\RR^{s-1}$ with $s>1$. The case $s=2$ gives rise to the 
so called quantum walls. In general, for $s>1$ 
the fields propagating in $E_i$ induce a non-trivial quantum field theory on $V$, 
which is no longer a point. The critical behavior of such kind of theories, with 
interactions localized exclusively on $V$, has been investigated in 
\cite{Fichera:2005fp}. On the other hand, phenomenological models 
with one edge and $s=4$, aiming to construct effective theories of 
the fundamental particle interactions, 
have been introduced in \cite{Randall:1999vf} and still attract much 
attention. In this case the vertex $V=\RR^3$ is interpreted as a 3-brane 
confining our universe. 

Concerning the applications to condensed matter physics, it is 
essential to construct quantum field theory models describing 
the conductance properties of quantum wires. It will be interesting 
in this respect to extend the bosonization procedure and the vertex 
algebra construction of \cite{Liguori:1997vd, Mintchev:2005rz} 
to star graphs with any number of edges. 
The case of general graphs with $N>1$ vertices, which are thus 
modeling more realistic quantum wires, represents a challenging problem 
as well. We are currently investigating some of these issues. 

\bigskip 
\bigskip

\noindent {\Large\bf Appendix}

\bigskip 
\bigskip

The orthogonality relation (\ref{chiort}) of the wave functions (\ref{chi}) is a 
consequence of the self-adjointness of the operator 
$-\prt_x^2$ on the graph $G$ with the boundary condition (\ref{bc}).  In order check this 
property directly, it is convenient to introduce the $T$-matrix defined by 
\be 
S(k) = \II_n + \ri \, T(k) \,  . 
\label{T1}
\ee
The explicit form of $T$, following from (\ref{cond1}, \ref{S1}, \ref{S2}), is  
\be 
T(k) =  (A+\ri kB )^{-1} 2\ri A=2\ri A^* (A^*+\ri kB^* )^{-1} \, .
\label{T2}
\ee
The integral over $x\in \RR_+$ in (\ref{chiort}) can be computed by means of 
the well-known identity 
\be 
-\ri \int_{0}^{+\infty} dx\, \e^{\ri k x} = \frac{1}{k+\ri \varepsilon} = 
\frac{1}{k} - \ri \pi \delta (k) \, , 
\label{a0}
\ee
the distribution $1/k$ being defined by means of the principal value prescription. 
One finds 
\be
\int_{0}^{+\infty} dx\, \chi_{\, i}^{*l}(x;k) \chi_l^j(x;p) = \delta_i^j\, 2\pi\, \delta (k-p)+
\T_i^j(k,p) \, , 
\label{a1}
\ee
where 
 \be 
 \T_i^j(k,p) = 
\frac{1}{p-k}[T_i^j(k)+T_i^j(-p)+\ri\, T_i^l(k)T_l^j(-p)]-\frac{1}{p+k}[T_i^j(k)-T_i^j(-p)] \, . 
\label{a2} 
\ee 
The final step is to prove that the matrix $\T(k,p)$ vanishes. Using (\ref{T2})  
one has in fact  
\begin{eqnarray}
T(k)+T(-p)&=& 4\ri A^* [AA^*+kpBB^*+\ri(k-p)AB^*]^{-1} A +           \\
          && 2p A^*  [AA^*+kpBB^*+\ri(k-p)AB^*]^{-1} B -  \nonumber\\
	  && 2kB^*   [AA^*+kpBB^*+\ri(k-p)AB^*]^{-1} A  \, ,   \nonumber
\end{eqnarray}
\begin{eqnarray}
T(k)-T(-p)&=& 2p A^* [AA^*+kpBB^*+\ri(k-p)AB^*]^{-1} B +           \\
          && 2k B^* [AA^*+kpBB^*+\ri(k-p)AB^*]^{-1} A  \, , \nonumber 
\end{eqnarray}
\be
\, \, \, 
T(k)T(-p)= -4 A^* [AA^*+kpBB^*+\ri(k-p)AB^*]^{-1} A \, ,
\ee
which allow to rewrite $\T(k,p)$ in the form 
\bea
\T(k,p) = 
\frac{4k}{p^2-k^2} \{p A^*[AA^*+kpBB^*+\ri(k-p)AB^*]^{-1} B - 
\nonumber \\
p B^*[AA^*+kpBB^*+\ri(k-p)AB^*]^{-1} A\}\, .  \, \,   
\label{a3}
\eea
The vanishing of (\ref{a3}) now follows from the condition (\ref{cond1}), which implies
\bea
[AA^*+kpBB^*+\ri(k-p)AB^*]^{-1}=(A^*+\ri kB^*)^{-1}(A-\ri pB)^{-1}=
\nonumber \\
(A^*-\ri pB^*)^{-1}(A+\ri kB)^{-1}\, . \, \, \,  \, \, 
\eea


\end{document}